%% file: BBeeck_revised.tex
\documentclass[mypaper,8pt,twoside]{CoAst}
\usepackage{epsf,graphicx,fancyhdr,sfmath}
\input{CoAst_jenam2008-layo}

\def\rfr{\smallskip\par\noindent
        \hangindent=7truemm
        \hangafter=1}

\begin{document}
\sf
\chapterCoAst{Towards a dynamical mass of a PG\,1159 star:
radial velocities and spectral analysis of SDSS\,J212531$-$010745}
{B.\,Beeck, S.\,Schuh, T.\,Nagel, and I.\,Traulsen}
\Authors{B.\,Beeck,$^{1}$ S.\,Schuh,$^{1}$ T.\,Nagel,$^{2}$ and I.\,Traulsen\,$^{1}$} 
\Address{
  $^1$ Institut f\"ur Astrophysik, Universit\"at G\"ottingen,
  Friedrich-Hund-Platz~1, 37077~G\"ottingen, Germany\\
  $^2$ Institut f\"ur Astronomie und Astrophysik, 
  Universit\"at T\"ubingen,
  Sand~1, 72076 T\"ubingen, Germany
}
\noindent
\begin{abstract}
The evolutionary scenarios which are commonly accepted for PG\,1159 stars are
mainly based on numerical simulations, which have to be tested and calibrated
with real objects with known stellar parameters. One of the most crucial but also quite uncertain parameters is the
stellar mass. PG\,1159 stars have masses between 0.5 and 0.8~M$_{\odot}$,
as derived from asteroseismic and spectroscopic determinations. Such mass
determinations are, however, themselves model-dependent. Moreover,
asteroseismically and spectroscopically determined masses deviate
systematically for PG\,1159 stars by up to 10\%.

SDSS~J212531.92-010745.9 is the first known PG\,1159 star in a close binary
with a late-main-sequence companion allowing a dynamical mass
determination. We have obtained 14 Calar Alto spectra of
SDSS~J212531.92$-$010745.9 covering the full orbital phase range. A radial velocity curve was extracted for both components. With co-added
phase-corrected spectra the spectral analysis of the PG\,1159 component was
refined. The irradiation of the companion by the PG\,1159 star is
modelled with \texttt{PHOENIX}, yielding constraints on radii, effective
temperature and separation of the system's components. The light curve
of SDSS~J212531.92$-$010745.9, obtained during three seasons of
photometry with the G\"ottingen 50\,cm and T\"ubingen 80\,cm telescopes, 
was modelled with both the \texttt{nightfall} and
\texttt{PHOEBE} programs. 
\end{abstract}

\Objects{SDSS~J212531.92-010745.9}

\section*{Extraction of the radial velocity curves}
In August 2007, 14 spectra of SDSS~J212531.92-010745.9 covering the total phase range were taken
with the TWIN spectrograph at the 3.5m telescope at Calar Alto Observatory (Alm{\'e}ria,
Spain). These show typical PG 1159 features together with the Balmer series of
hydrogen in emission (plus other emission lines), already interpreted as signature of
an irradiated close companion by Nagel et al.\ (2006) from
an SDSS spectrum. The Calar Alto spectra cover wavelengths from 3800\,\AA\ up
to about 7000\,\AA . At 5000\,\AA\ they have a resolution of $R=4170$ and a
SNR ranging from 4 to 13. The spectra were reduced and normalized to their 
continua. To determine the
radial velocity (RV) curve of the secondary of the system a Gaussian was
fitted to seven H Balmer
lines (H\,$\alpha$, H\,$\beta$,~..., H\,$\eta$, emission line height variable
with phase) to locate the line
centres. To evade uncertainties in the wavelength calibration and intrinsic wavelength dependence of the RV,
a sine was fitted to the RV curve separately for each line. The
zero points of the fits were shifted to the mean zero point $\langle
v_0\rangle = 81.5\,\mathrm{km\,s}^{-1}$ with an uncertainty of $\pm 9.4\,\mathrm{km\,s}^{-1}$. The weighted mean of the shifted RVs
was calculated and a sine was fitted to the resulting mean RVs yielding a secondary RV
amplitude of $K_2=(113.0 \pm 3.0)\,\mathrm{km\,s}^{-1}$.
The primary RV curve was deduced from the cross correlation of a small section of
the spectra containing two narrow C{\footnotesize\,IV} absorption lines at
5801/5812\,\AA, which were unblended with the strong emission lines of the companion. The accuracy of the RV values obtained by this method was
estimated to amount to $\pm 30\,\mathrm{km\,s}^{-1}$ and was rescaled with the SNR
of the individual spectra. Again a sine was fitted to the RVs. The amplitude of this fit
is $K_1=(94.3\pm 15.0)\,\mathrm{km\,s}^{-1}$ and its zero point is
at $v_{0,\mathrm{PG\,1159}}=(92\pm 33)\,\mathrm{km\,s}^{-1}$, slightly redshifted with respect to the zero
point of the RV curve of the secondary. Although this redshift is insignificant, it raises the
hope to detect a gravitational redshift for the PG\,1159 component of SDSS~J212531.92$-$010745.9 with
high-resolution spectroscopy in subsequent studies. The RV curves are shown in
Fig.~\ref{fig:beeck_fig1}.
\figureCoAst{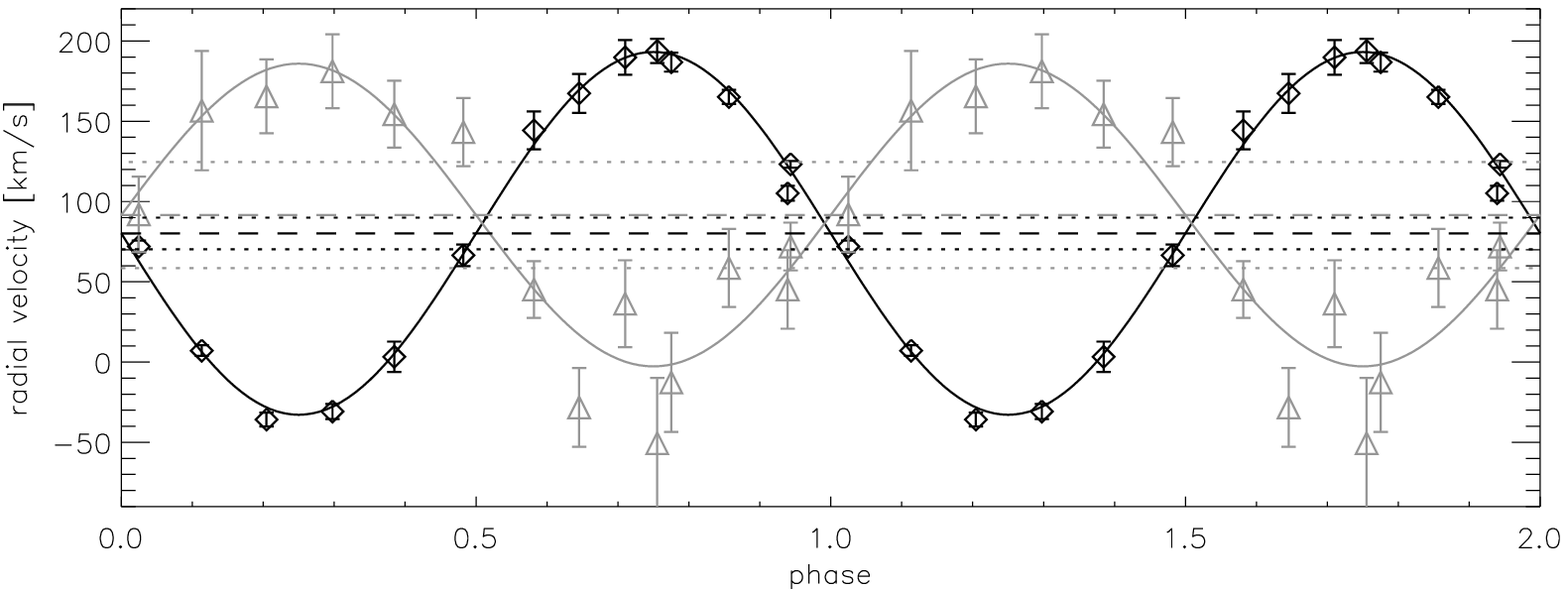}{Radial
  velocities (RVs) derived from the Calar Alto spectra. The grey symbols and error
  bars signify the RVs of the PG\,1159 component together with the sine fit
  (solid grey line; amplitude $K_1=(94.3\pm 15.0)\mathrm{km\,s}^{-1}$), the zero point of which is shown as dashed line. The black
  symbols signify the RVs of the secondary, again with the sine fit (solid
  black line; amplitude $K_2=(113.0\pm 3.0)\mathrm{km\,s}^{-1}$) and its zero point (dashed black line). The dotted lines are the
  uncertainties of the zero points.}
{fig:beeck_fig1}{t}{clip,angle=0,width=0.85\textwidth 
} 
\section*{Spectral analysis and light curve modelling}
Using the RV curves obtained, the spectra were shifted to zero redshift for
the individual components of the system to determine a mean and median of
both the primary and secondary spectrum. A grid of model spectra was computed with the
non-LTE atmosphere modelling package \texttt{NGRT} (Werner et al.\ 2003) and fitted to the median
primary spectrum. The best fit so far is obtained for a model with
$T_{\mathrm{eff}}=72\,500\,\mathrm{K}$, $\log g = 7.20$ and abundances (by number) of
C/He=0.07, N/He=0.01, O/He=0.01, implying a mass fraction
(He/C/N/O)=(0.78/0.16/0.03/0.03), typical for PG\,1159
stars -- see e.\,g.~Werner \& Herwig (2006). Typical errors for this kind of
analysis are $\pm 10\,000\,\mathrm{K}$ for $T_{\mathrm{eff}}$ and $\pm 0.3$
for $\log g$ (Ziegler et al., these proceedings). The temperature obtained is
significantly lower than the one found by Nagel et al.\ (2006), who obtained the preliminary values
$T_{\mathrm{eff}}=90\,000\,\mathrm{K}$ and $\log g = 7.60$, and implies
that SDSS~J212531.92$-$010745.9 is one of the coolest PG\,1159 stars.

To get the stellar parameters of the irradiated secondary, a second model grid is being
calculated using \texttt{PHOENIX} (Hauschildt et al.\ 1997). Up to now, no quantitative result has been
obtained, but the Balmer series of hydrogen in emission is reproduced by the
first models calculated.

Three seasons of photometry are available for SDSS~J212531.92$-$010745.9. The system shows flux variations with a peak-to-peak
amplitude of about 0.7~mag and a period of about 6.96~h. The
ephemeris could be determined to a high accuracy. 
The light curve profile obtained shows no eclipse and is currently being fitted with the binary
modelling programs \texttt{nightfall} and \texttt{PHOEBE} (Pr{\v{s}}a \& Zwitter 2005). This light curve
modelling possibly constrains the inclination of the system which is needed to
deduce the mass of the components from the RVs. Together with the 
constraints of an optimized spectral analysis (especially with 
\texttt{PHOENIX}) this will give a possible mass range for the PG\,1159 
component of SDSS~J212531.92-010745.9 (work in progress: Beeck 2009,
Schuh et al.\ 2009, and references therein).\\

\acknowledgments{The spectroscopy is based on service observations collected
  by J.~Aceituno and U.~Thiele
  at the Centro Astron{\'o}mico Hispano
  Alem{\'a}n, operated jointly by the Max-Planck
  Institut f\"ur Astronomie and the Instituto de Astrof{\'i}sica de
  Andaluc{\'i}a.
  Special thanks to B.~G\"ansicke and M.~Schreiber who first directed
  our attention to this unique object, and to all observers for
  the photometric observations.
  We also thank the Astronomische Gesellschaft as well as the
  conference sponsors and in particular HELAS 
  for financially supporting the poster presentation at JENAM 2008
  through travel grants to B.B.\ and S.S.}

\References{
\rfr Beeck, B.\ 2009, Diploma thesis, University of G\"ottingen, in prep.
\rfr Hauschildt, P., Baron, E., \& Allard, F.\ 1997, ApJ, 490, 803
\rfr Nagel, T., Schuh, S., Kusterer, D.-J., et~al.\ 2006, {A\&A}, {448}, L25
\rfr Pr{\v{s}}a, A., \& Zwitter, T.\ 2005, ApJ, 628, 426
\rfr Schuh, S., Beeck, B., \& Nagel, T.\ 2009,
in {''White Dwarfs''}, {J.\ Phys.: Conf.\ Ser.}, in press, arXiv:0812.4860
\rfr Werner, K., Deetjen, J.~L., Dreizler, S., et al.\ 2003,  
in {''White Dwarfs''}, {NATO ASIB Proc.}, {105}, 117
\rfr Werner, K., \& Herwig, F.\ 2006, {PASP}, {118}, 183
}
\end{document}

%% file: CoAst_jenam2008-layo.tex
\renewcommand{\chaptermark}[1]{\markboth{#1}{}}

\newcommand{\kopf}{\small\itshape Comm.\ in Asteroseismology, N$^{\textsf{\underline{o}}}$ 159, 2009\\
Proceedings of the JENAM 2008 Symposium N$^{\textsf{\underline{o}}}$~4:
Asteroseismology and Stellar Evolution}

\newcommand{\Authors}[1]{\begin{center}\normalsize\bf\sf #1 \end{center}}

\renewcommand{\author}[1]{\begin{center}\normalsize\bf\sf #1 \end{center}}
\newcommand{\Address}[1]{\begin{center}\small\sf #1 \end{center}}

\newcommand{\Objects}[1]{{\vspace{3mm}\small \noindent Individual Objects: }\small\sf \hangindent=27truemm \hangafter=1 #1 \normalsize}

\renewenvironment{abstract}{\section*{Abstract}\normalsize\sf}{}
\newcommand{\References}[1]{\begin{flushleft}{\large References\\}\vspace*{2mm}\small #1 \end{flushleft}}

\newcommand{\chapterCoAst}[2]{\chapter[\sf\normalsize #1\\ \footnotesize \hspace*{5mm}by #2 \sf\normalsize][]{#1\\}\rhead[\fancyplain{}{\sf\footnotesize \center{#1}}]{\fancyplain{}{\sffamily\thepage}}\lhead[\fancyplain{\kopf}{\sffamily\thepage}]{\fancyplain{\kopf}{\sf\footnotesize \center{#2}}}}

\newcommand{\figureCoAst}[5]{\begin{figure}[#4]
\centering
\includegraphics*[#5]{#1}
\caption{#2}
\label{#3}
\end{figure}}

\renewcommand{\chaptername}{}

\newcommand{\acknowledgments}[1]{\vspace*{5mm}\noindent\textbf{Acknowledgments.} #1}